\documentclass[aps,preprint,preprintnumbers,nofootinbib,showpacs,superscriptaddress]{revtex4}
\usepackage{graphicx,color,amsmath}
\begin{document}
\def\be{\begin{eqnarray}}
\def\en{\end{eqnarray}}
\def\non{\nonumber}
\def\la{\langle}
\def\ra{\rangle}
\def\B{{\cal B}}
\def\ov{\overline}
\def\lsim{ {\ \lower-1.2pt\vbox{\hbox{\rlap{$<$}\lower5pt\vbox{\hbox{$\sim$}
}}}\ } }
\def\gsim{ {\ \lower-1.2pt\vbox{\hbox{\rlap{$>$}\lower5pt\vbox{\hbox{$\sim$}
}}}\ } }

\font\el=cmbx10 scaled \magstep2{\obeylines\hfill June, 2009}
\vskip 1.5 cm
\title{Study of $\bar B\to \Lambda_c\bar \Lambda_c$ and $\bar B\to \Lambda_c\bar \Lambda_c \bar K$}
\author{Hai-Yang  Cheng}
\affiliation{Institute of Physics, Academia Sinica, Taipei, Taiwan 115, R.O.C.}
\author{Chun-Khiang Chua}
\affiliation{Department of Physics, Chung Yuan Christian University,\\ Chung-Li, Taiwan 320, R.O.C.}
\author{Yu-Kuo Hsiao}
\affiliation{Institute of Physics, Academia Sinica, Taipei, Taiwan 115, R.O.C.}
\date{\today}
\begin{abstract}
We study the doubly charmful two-body and three-body baryonic $B$
decays $\bar B\to \Lambda_c^+\bar \Lambda_c^-$ and $\bar B\to
\Lambda_c^+\bar \Lambda_c^- \bar K$. As pointed out before, a
naive estimate of the branching ratio ${\cal O}(10^{-8})$ for the
latter decay  is too small by three to four orders of magnitude
compared to experiment. Previously, it has been shown that a large
enhancement for the $\Lambda_c^+\bar\Lambda_c^-\bar K$ production
can occur due to a charmonium-like resonance (e.g. $X(4630)$
discovered by Belle) with a mass near the $\Lambda_c\bar\Lambda_c$
threshold. Motivated by the BaBar's observation of a resonance in
the $\Lambda_c \bar K$ system with a mass of order 2930 MeV,  we
study in this work the contribution to $\bar
B\to\Lambda_c^+\bar\Lambda_c^-\bar K$ from the intermediate state
$\Xi_c(2980)$  which is postulated to be a first positive-parity
excited $D$-wave charmed baryon state.  Assuming that a soft $q\bar q$ quark pair is produced through the $\sigma$ and $\pi$ meson
exchanges in the configuration for $\bar B\to
\Xi_c(2980)\bar\Lambda_c$ and $\Lambda_c\bar\Lambda_c$, it is
found that  branching ratios of  $\bar B\to \Lambda_c^+\bar
\Lambda_c^- \bar K$ and $\bar B\to \Lambda_c^+\bar \Lambda_c^-$
are of order $3.5\times 10^{-4}$ and $5\times 10^{-5}$, respectively, in agreement with experiment except that the prediction for the $\Lambda_c\bar\Lambda_c K^-$ is slightly smaller. In conjunction with our previous analysis, we conclude that the enormously large rate of $\bar B\to \Lambda_c^+\bar \Lambda_c^-\bar K$ arises from the resonances $\Xi_c(2980)$ and $X(4630)$.

\end{abstract}


\maketitle
\newpage
\section{introduction}
There are several unique features in baryonic $B$ decays. First,  a
peak near the threshold area of the dibaryon invariant mass
spectrum has been observed in many baryonic $B$ decays. Second,
three-body decays usually have rates  larger than their two-body
counterparts; that is, $\B(B\to {\bf B\bar B'}M)\gg \B(B\to {\bf
B\bar B'})$. This phenomenon can be understood in terms of the
threshold effect, namely, the invariant mass of the dibaryon is
preferred to be close to the threshold. The configuration of the
two-body decay $B\to{\bf B\bar B'}$ is not favorable since its
invariant mass is $m_B$. In $B\to {\bf B \bar B'} M$ decays, the
effective mass of the baryon pair is reduced as the emitted meson
can carry away much energy. The low mass threshold effect can be
understood in terms of a simple short-distance picture
\cite{Suzuki}. For singly charmful baryonic $B$ decays,
experimentally we have $\B(B^-\to \Lambda_c^+\bar p
\pi^-\pi^0)>\B(B^-\to \Lambda_c^+\bar p \pi^-)>\B(\bar B^0\to
\Lambda_c^+\bar p)$  \cite{exampleC,LambdaCbarp} and $\B(\bar B^0\to
D^{(*)+}\pi^- p\bar p)>\B(\bar B^0\to D^{(*)0} p\bar p)$
\cite{ppD}. Therefore, we have a pattern like
\begin{eqnarray}
&&\B(\bar B\to {\bf B_{(c)}\bar B'}MM')> \B(\bar B\to {\bf B_{(c)}\bar B'}M)\gg \B(\bar B\to {\bf B_{(c)}\bar B'})\,,
\end{eqnarray}
where $\bf B_{c}$ denotes a charmed baryon.

The experimental measurements for doubly charmful $B$ decays are
summarized in Table \ref{tab:2bodycharm}. For $B\to \Xi_c\bar
\Lambda_c$ decays, we extract their branching ratios using
$\B(\Xi_c^0\to \Xi^-\pi^+)=1.3\%$ and $\B(\Xi_c^+\to \Xi^-
\pi^+\pi^+)=3.9\%$ \cite{ChengChuaTsi,Cheng&Tseng}, respectively,
\begin{eqnarray}
\B(B^-\to \Xi_c^0\bar \Lambda_c^-)&=&(2.0\pm 0.6^{+1.1}_{-0.5})\times 10^{-3}~{\rm (average~of ~BaBar~and~Belle)}\,,\nonumber\\
\B(\bar B^0\to \Xi_c^+\bar \Lambda_c^-)&=&(3.8\pm 3.1^{+8.7}_{-2.4})\times 10^{-4}<1.4\times 10^{-3}~{\rm (BaBar)}, \nonumber \\
                                      &=&(2.4\pm 1.2^{+5.3}_{-1.5})\times 10^{-3}~{\rm (Belle)}\,,
\end{eqnarray}
where the second errors originate from the uncertainties in
$\B(\Xi_c^0\to \Xi^-\pi^+)$ ranging from 0.83\% to 1.74\% and
$\B(\Xi_c^+\to \Xi^- \pi^+\pi^+)$ from 1.2\% to 10.1\%
\cite{Korner}. Theoretically, it is expected that the
charged and neutral $B$ decays to $\Xi_c\bar\Lambda_c$ should have
similar rates. Experimentally, this feature should be tested by
the forthcoming measurements. Since $\B(\bar B\to \Lambda_c\bar
p)\approx 2\times 10^{-5}$ \cite{LambdaCbarp}, we have another pattern
\begin{eqnarray}
&&\B(\bar B\to {\bf B_c\bar B'_c})\sim 10^{-3}\gg \B(\bar B\to {\bf B_c\bar B'})\sim 10^{-5}\gg \B(\bar B\to {\bf B\bar B'})\lesssim 10^{-7}\;\;
\end{eqnarray}
for two-body baryonic $B$ decays.

\begin{table}[t]
\caption{Branching ratios  of doubly charmful two-body (in units
of $10^{-5}$) and three-body (in units of $10^{-4}$) baryonic $B$
decays. } \label{tab:2bodycharm}
\begin{ruledtabular}
\begin{tabular}{l l l  }
 Decay & BaBar \cite{LambdaCLambdaCK} & Belle \cite{LambdaCLambdaC,LambdaCLambdaCK0} \\ \hline
 $B^-\to\Xi_c^0(\to\Xi^-\pi^+)\bar\Lambda_c^-$ & $2.08\pm0.65\pm0.29\pm0.54$ &
 $4.8^{+1.0}_{-0.9}\pm1.1\pm1.2$  \\
 $\overline B^0\to\Xi_c^+(\to\Xi^-\pi^+\pi^+)\bar\Lambda_c^-$ & $1.50\pm1.07\pm0.20\pm0.39<5.6$ &
 $9.3^{+3.7}_{-2.8}\pm1.9\pm2.4$  \\
 $\overline B^0\to\Lambda_c^+\bar\Lambda_c^-$ & & $2.2^{+2.2}_{-1.6}\pm1.3<6.2$ \\
 \hline
 $B^-\to \Lambda_c^+\bar\Lambda_c^-K^-$ & $11.4\pm1.5\pm1.7\pm6.0$ &
 $6.5^{+1.0}_{-0.9}\pm1.1\pm3.4$  \\
 $\overline B^0\to \Lambda_c^+\bar\Lambda_c^-\overline K^0$ & $3.8\pm3.1\pm0.5\pm2.0<15$ &
 $7.9^{+2.9}_{-2.3}\pm1.2\pm4.1$  \\
\end{tabular}
\end{ruledtabular}
\end{table}

Since the doubly charmed
baryonic decay $\bar B\to \Xi_c\bar\Lambda_c$ proceeds via $b\to cs\bar
c$, while $\bar B\to \Lambda_c\bar p$ via a $b\to cd\bar u$ quark
transition, the CKM mixing angles for them are the same in
magnitude but opposite in sign. One may wonder why the ${\bf B_c\bar B'_c}$ mode has a rate two orders of magnitude larger than
${\bf B_c\bar B}$.
According to the conjecture made by Hou and Soni \cite{Hou&Soni}, one has to reduce the energy
release and  at the same time allow for baryonic ingredients to be
present in the final state in order to
have larger baryonic $B$ decays. Hence, it
is expected that
 \be
 \Gamma(B\to {\bf B_1\bar B_2})=|{\rm CKM}|^2/f({\rm energy~release})=|{\rm CKM}|^2/(Q~{\rm value}),
 \en
where CKM stands for the relevant CKM angles. For charmful modes,
one will expect
 \be
 \B(\ov B^0\to\Lambda_c^+\bar p)=|V_{ud}/V_{cs}|^2\B(\ov
 B^0\to\Xi_c^+\bar\Lambda_c^-)({\rm dynamical~suppression}),
 \en
where the dynamical suppression arises from the larger energy
release in $\Lambda_c^+\bar p$ than in $\Xi_c\bar\Lambda_c$. This
is because no hard gluon is needed to produce the energetic
$\Xi_c\bar\Lambda_c$ pair in the latter decay, while two hard
gluons are needed for the former process \cite{ChengChuaTsi}.
Therefore, $\Lambda_c\bar p$ is suppressed relative to
$\Xi_c\bar\Lambda_c$ due to a dynamical suppression from ${\cal
O}(\alpha_s^4)\sim 10^{-2}$. These qualitative statements have
been confirmed by the realistic calculations of $\bar B\to
\Xi_c\bar\Lambda_c$ in \cite{ChengChuaTsi} and $\bar B\to
\Lambda_c\bar p$ in \cite{He}.


For $\bar B^0\to \Lambda_c^+\bar \Lambda_c^-$, we expect a
branching ratio of order $10^{-5}$ from the estimate of $\B(\bar
B^0\to \Lambda_c^+\bar \Lambda_c^-)\simeq
|V_{cd}/V_{cs}|^2\,\B(\bar B^0\to \Xi_c^+\bar \Lambda_c^-)$ and
from $|V_{cd}/V_{ud}|^2\B(\bar B^0\to \Lambda_c\bar p)/({\rm
dynamical~suppression})$. Hence, the expected branching ratio
obtained from the naive extrapolation from $\B(\bar B^0\to
\Xi_c^+\bar \Lambda_c^-)$ and from $\B(\bar B^0\to \Lambda_c\bar
p)$ is in accordance with experiment.

\begin{figure}[t]
\centering
\includegraphics[width=2.0in]{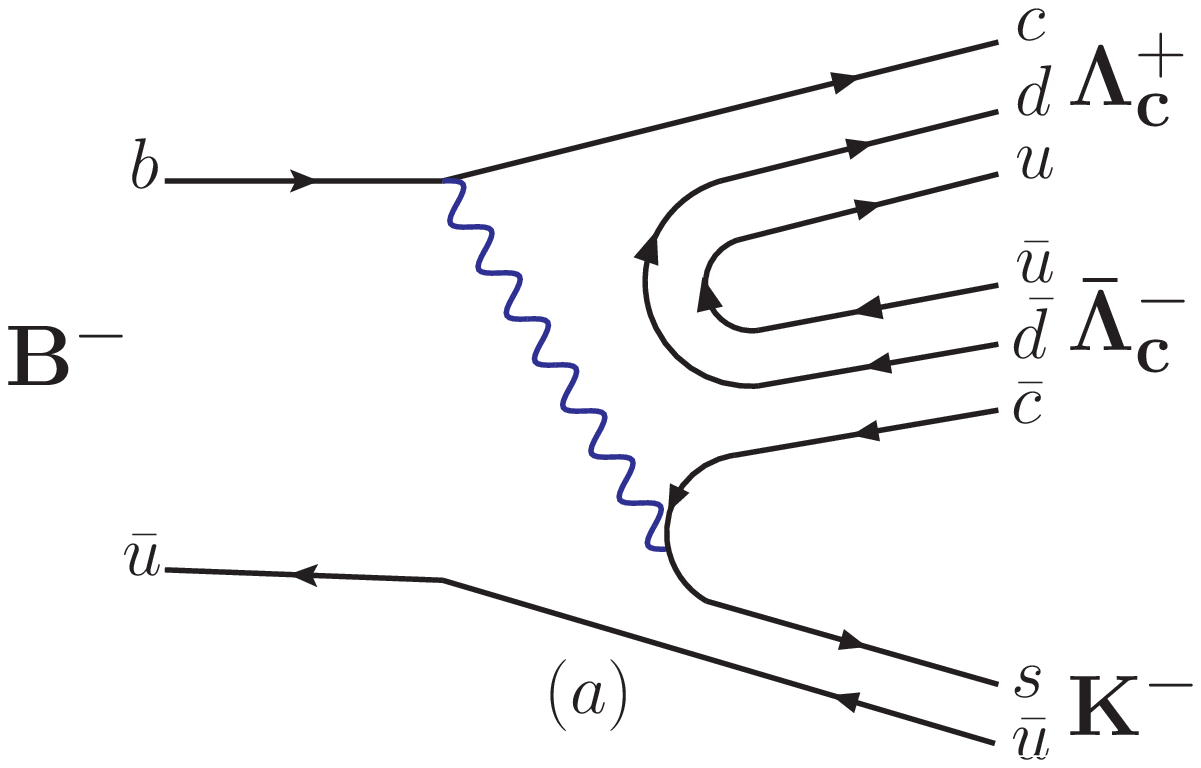}
\includegraphics[width=2.0in]{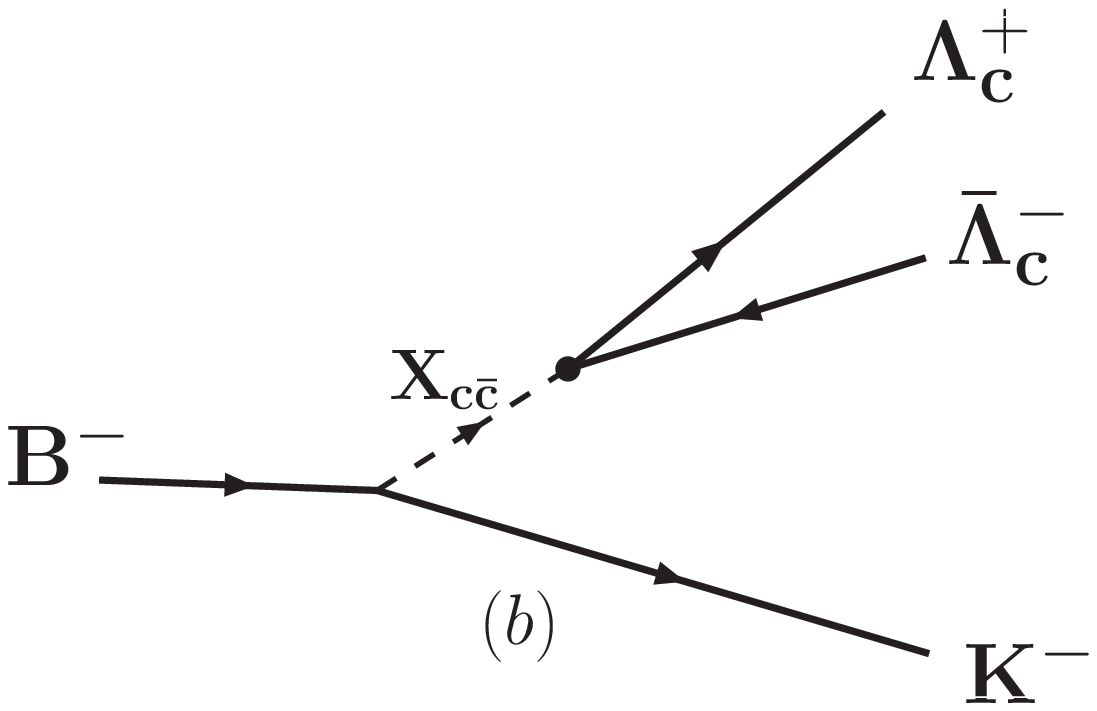}
\includegraphics[width=2.0in]{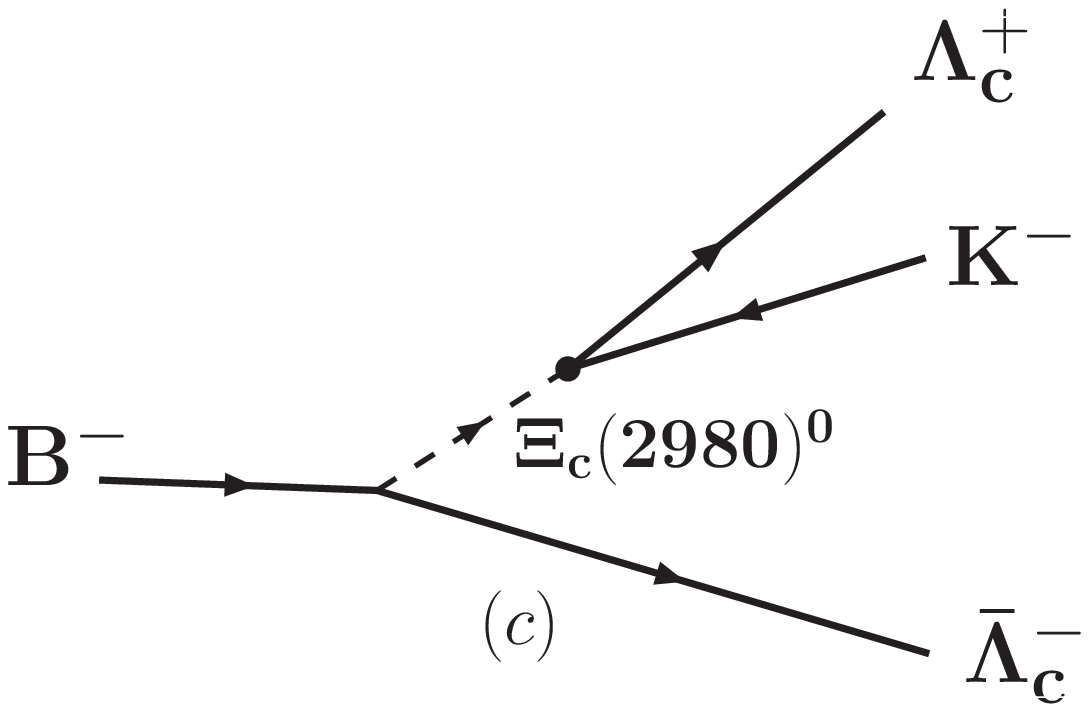}
\caption{$B^-\to \Lambda^+\bar \Lambda^- K^-$ as proceeding through
(a) the internal W-emission diagram, (b) the dominant charmonium-like resonance $X_{c\bar c}$,
and (c) the resonant state of D-wave $\Xi_c(2980)^0$. The blob in (b) and (c) shows where the strong decays take place.} \label{fig1}
\end{figure}

The three-body doubly charmed baryonic decay
$\bar B\to\Lambda_c\bar\Lambda_c\bar K$ has been observed at $B$ factories
with the branching ratio of order $(10^{-3}-10^{-4})$
\cite{LambdaCLambdaCK,LambdaCLambdaCK0}. Since this mode is
color-suppressed and its phase space is highly suppressed, the
naive estimate of $\B(\bar B\to\Lambda_c\bar\Lambda_c\bar K)\sim
10^{-8}$ from Fig. \ref{fig1}(a) is too small by three to four orders of magnitude
compared to experiment. It was originally conjectured in
\cite{ChengChuaTsi} that the great suppression for the
$\Lambda_c^+\bar\Lambda_c^-\bar K$ production can be alleviated
provided that there exists a hidden charm bound state $X_{c\bar c}$ with a mass
near the $\Lambda_c\bar\Lambda_c$ threshold [see Fig. \ref{fig1}(b)], of order 4.6 $\sim$
4.7 GeV. This possibility is motivated by the observation of many
new charmonium-like resonances with masses around 4 GeV starting
with $X(3872)$  and so far ending with $Z(4430)$ by BaBar and
Belle. This new state that couples strongly to the charmed baryon
pair can be searched for in $B$ decays and in $p\bar p$ and
$e^+e^-$ collisions by studying the mass spectrum of $D^{(*)}\ov
D^{(*)}$ or $\Lambda_c\bar\Lambda_c$. However, an initial
investigation of the $\Lambda_c\bar\Lambda_c$ spectrum in the
$\bar B\to\Lambda_c\bar\Lambda_c\bar K$ decays by Belle did not reveal any
new resonance with a mass near the $\Lambda_c\bar\Lambda_c$
threshold (see Fig. 3 in version 2 of \cite{LambdaCLambdaCK0}).
Nevertheless, the situation was  dramatically changed recently.
Using initial-state radiation, Belle has reported a near-threshold
enhancement in the $e^+e^-\to \Lambda_c^+\Lambda_c^-$ exclusive
cross section \cite{BelleLambdac}. With an assumption of a
resonance origin for the observed peak, called the $X(4630)$,
Belle obtained $m=4634^{+8+5}_{-7-8}$ MeV and
$\Gamma=92^{+40+10}_{-24-21}$ MeV. Interestingly, these values are
consistent within errors with the mass and width of the $Y(4660)$
with $J^{PC}=1^{--}$ found in $\psi(2S)\pi\pi$ decays
\cite{BelleY4660}.

Other possibilities for the enhancement of $\Lambda_c\bar\Lambda_c \bar K$
rates include final-state interactions and $\Lambda_c\bar K$
resonances. For the first possibility, the weak decay $\bar B\to
D^{(*)}\bar D_s^{(*)}$ followed by the rescattering of
$D^{(*)}\bar D_s^{(*)}$ to $\Lambda_c\bar\Lambda_c\bar K$ has been
considered in  \cite{Chen}. For the second possibility, BaBar has
recently studied possible intermediate states in
$\bar B\to\Lambda_c\bar\Lambda_c\bar K$ and found a resonance in the
$\Lambda_c \bar K$ invariant mass distribution \cite{LambdaCLambdaCK}
\begin{eqnarray}
m=2931\pm 3\pm 5\; \text{MeV}\,,\;\;\Gamma= 36\pm 7\pm 11\;\text{MeV}\,.
\end{eqnarray}
This could be interpreted as a single $\Xi_c^0$ resonance. An
examination of the $\Xi_c$ spectroscopy  suggests that this
resonance can be identified with $\Xi_c(2980)$ \cite{pdg}
\begin{eqnarray} \label{width}
\Xi_c(2980)^+: && m_{\Xi'_c}=2974\pm 5\; \text{MeV}\,,\;\;\Gamma= 33\pm8\;\text{MeV}\,, \nonumber \\
\Xi_c(2980)^0: && m_{\Xi'_c}=2974\pm 4\; \text{MeV}\,,\;\;\Gamma= 31\pm11\;\text{MeV}\,.
\end{eqnarray}

In this work, we shall consider the $\Lambda_c \bar K$ resonant
contribution to $\bar B\to\Lambda_c\bar\Lambda_c\bar K$ from
$\Xi_c(2980)$ to see if it can lead to the anomalously large rate
for this decay mode (Fig. \ref{fig1}(c)). Besides, we also examine $\bar B^0\to
\Lambda_c^+\bar \Lambda_c^-$ to give a concrete prediction. This
paper is organized as follows. The formalism is given in Sec. II
followed by a numerical analysis. We then give a discussion on
physical results and conclude the paper in Sec. IV.

\section{formalism}

\begin{figure}[t]
\centering
\includegraphics[width=2.0in]{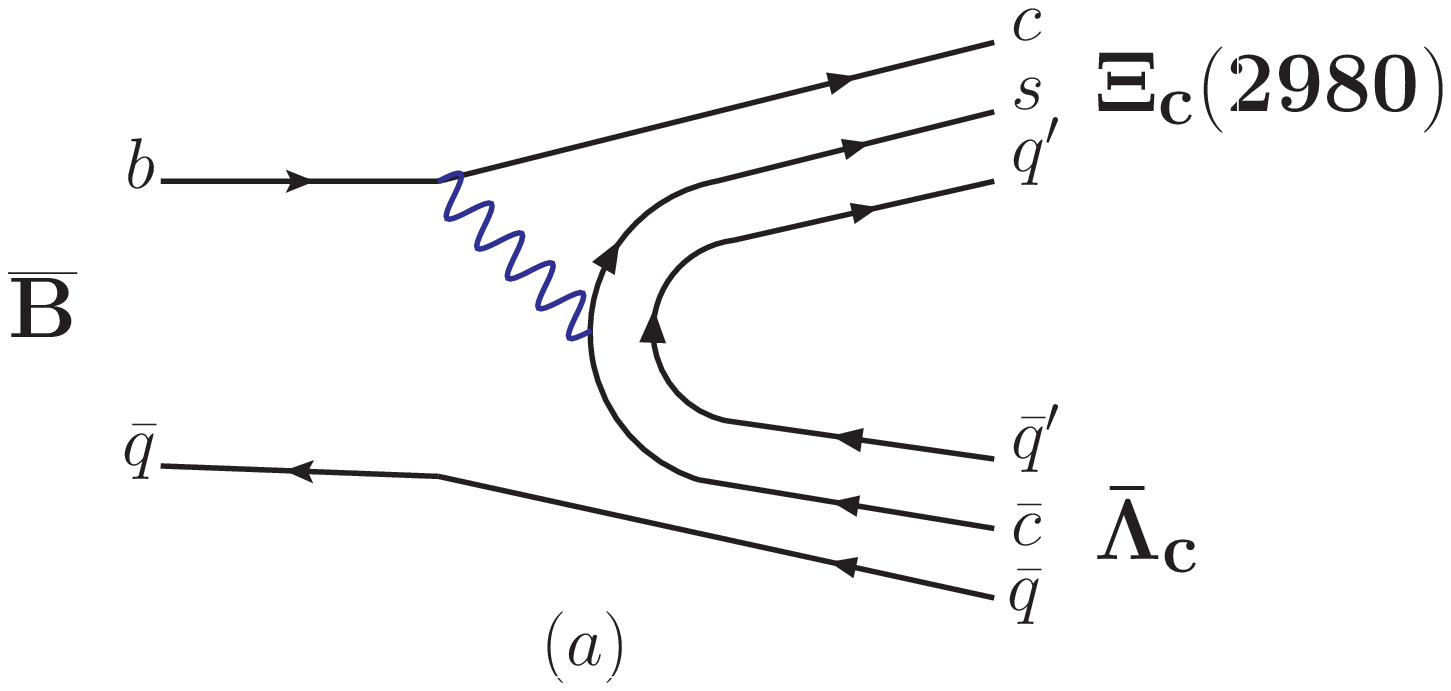}
\includegraphics[width=2.0in]{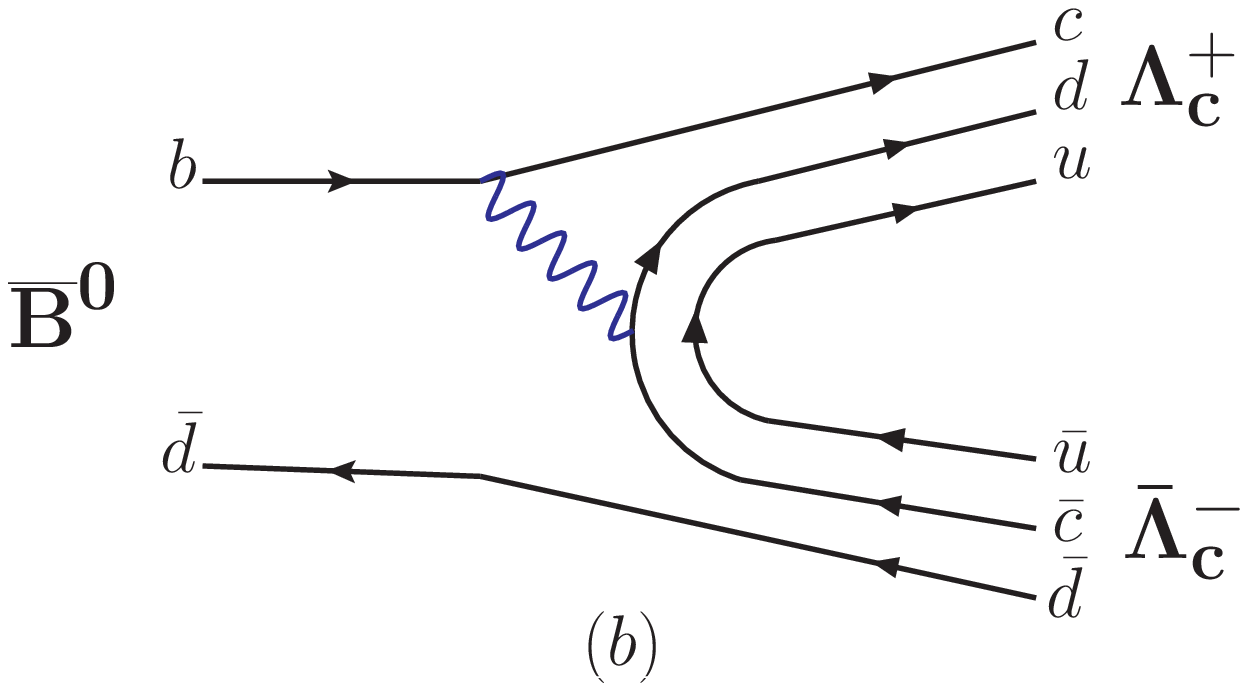}
\caption{(a) $\bar B\to \Xi_c(2980)\bar \Lambda_c$ and (b) $\bar B^0\to \Lambda_c^+\bar \Lambda_c^-$ as proceeding via internal W-emission diagrams.
In (a), $qq'=du$ and $ud$ for $B^-$ and $\bar B^0$ decays, respectively.}\label{fig2}
\end{figure}

\begin{figure}[t]
\centering
\includegraphics[width=4in]{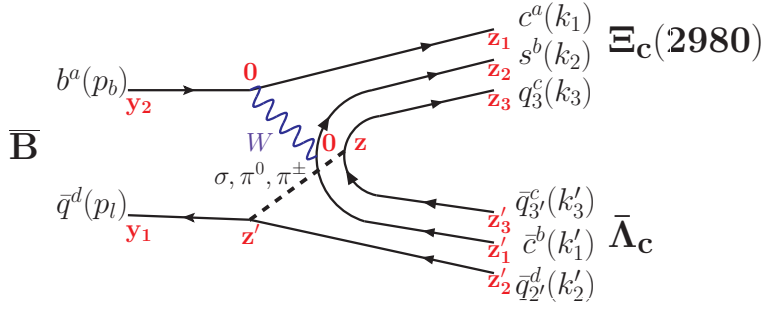}
\caption{$\bar B\to \Xi_c(2980)\bar \Lambda_c$, where $qq_3
q_{3'}q_2=uddu(udud)$ for $\sigma,\,\pi^0$ ($\pi^\pm$) exchange in
$B^-$ decays, and $qq_3 q_{3'}q_2=duud(dudu)$ for $\sigma,\,\pi^0$
($\pi^\pm$) exchange in $\bar B^0$ decays. }\label{charm}
\end{figure}

The Cabibbo-allowed two-body doubly charmed baryonic $B$ decays
$\bar B \to\Xi_c(2980)\bar\Lambda_c$ and Cabibbo-suppressed decay $B\to
\Lambda_c\bar\Lambda_c$ receive contributions from the internal
$W$-emission (Fig. \ref{fig2}) and weak annihilation. The
latter contribution can be safely neglected as it is not only
quark-mixing but also helicity suppressed. As mentioned in the
Introduction, we shall consider the $\Lambda_c \bar K$ resonant
contribution to $\bar B\to\Lambda_c\bar\Lambda_c \bar K$ from
$\Xi_c(2980)$ using the narrow width approximation
\begin{eqnarray}
\B(\bar B\to\Lambda_c\bar\Lambda_c \bar K)=\B(\bar
B\to\Xi_c(2980)\bar\Lambda_c)\B(\Xi_c(2980)\to\Lambda_c \bar K).
\end{eqnarray}
Since for an energetic charmed baryon its momentum
is carried mostly by the charmed quark, the two-body doubly charmful baryonic $B$ decays can proceed without a hard gluon. In other words, the $q\bar q$ pair (e.g. $q'\bar q'$ in Fig. \ref{fig2}(a) and $u\bar u$ in Fig. \ref{fig2}(b)) is likely produced  from the
vacuum via the soft nonperturbative interactions so that it
carries the vacuum quantum numbers $^3P_0$. Following
\cite{ChengChuaTsi}, we shall consider the possibility that the
$q\bar q$ pair is produced via a light meson exchange. The $q\bar
q$ pair created from soft nonperturbative interactions tends to be
soft.  To be specific, we assume the exchange of the $\sigma$,
$\pi^0$ and $\pi^-$ between the soft $q\bar q$ quark pair and the
spectator as shown in Fig.~\ref{charm}. It should be stressed that Fig. 3 here differs from Fig. 5 of \cite{ChengChuaTsi} as $\Xi_c$ in the latter is a ground-state $S$-wave cascade charmed baryon, while $\Xi_c(2980)$ in the former is an excited charmed baryon. Hence, a repeat of the analysis in \cite{ChengChuaTsi} will not provide any information on $B \to \Xi_c(2980)\bar\Lambda_c$.

To obtain the amplitudes of $B^-\to \Xi_c(2980)^0 \Lambda_c^-$ and
$\bar B^0\to \Lambda_c^+\bar \Lambda_c^-$, we start from the
short-distance effective Hamiltonian  given by
\begin{eqnarray}\label{Heff}
{\cal H}_{\rm eff}&=&\frac{G_F}{\sqrt 2}V_{cb}V_{cq}^*(c_1 O_1+c_2 O_2)\,,
\end{eqnarray}
where $O_1=(\bar c b)(\bar q c)$ and $O_2=(\bar c c)(\bar q b)$
with $q=s$ for $\Xi_c(2980)^0$, $q=d$ for $\Lambda_c^+$ and $(\bar
q q')\equiv\bar q\gamma_\mu(1-\gamma_5)q'$. We shall use the Wilson
coefficients $c_1=1.169$ and $c_2=-0.367$. The Lagrangian for
meson-quark interactions reads
\begin{eqnarray}\label{Lan}
{\cal L}_{\sigma qq}&=&g_\sigma(\bar u u+ \bar d d)\sigma\,,\nonumber\\
{\cal L}_{\pi^0 qq}&=&g_{\pi^0}(\bar u i\gamma_5 u-\bar d i\gamma_5 d)\pi^0\,,\nonumber\\
{\cal L}_{\pi^\pm qq}&=&g_{\pi}(\bar u i\gamma_5 d \pi^+ +\bar d i\gamma_5 u\pi^-)\,,
\end{eqnarray}
where $g_i$ ($i=\sigma,\;\pi^0,\;\pi$) is the coupling constant,
and $g_{\pi}=\sqrt 2 g_{\pi^0}$ from isospin symmetry. The
amplitude of $B^-\to \Xi_c(2980)^0\bar \Lambda_c^-$ in Fig.
\ref{charm} thus has the form
\begin{eqnarray}\label{amptotal}
{\cal A}={\cal A}_\sigma+{\cal A}_{\pi^0}+{\cal A}_{\pi^\pm}.
\end{eqnarray}
In the case of $\sigma$ exchange, the amplitude reads
\begin{eqnarray}\label{Asigma}
i{\cal A}_\sigma &=&\frac{G_F}{\sqrt 2}V_{cb}V_{cs}^*(c_1-c_2)\int d^4 z d^4 z' (ig_\sigma)^2 \langle \sigma(z)\sigma(z')\rangle
(-1)\Gamma_{\alpha\rho}\Gamma_{\beta\delta}\Gamma^\sigma_{\gamma\gamma'}\Gamma^\sigma_{\eta\eta'}\nonumber\\
 &\times& \langle\Xi_c(2980)^0|\bar c_\alpha^a(0)\bar s_\beta^b(0)\bar d_\gamma^c(z)|0\rangle
          \langle\overline{\Lambda}^-_c|c_\delta^b(0) u_{\eta'}^d(z')d_{\gamma'}^c(z) |0\rangle
          \langle 0|\bar u_\eta^d(z) b^a_\rho(0)
|B^-\rangle\,,
\end{eqnarray}
with the Latin superscripts denoting the color indices, the Greek subscripts the Dirac indices, and
$z_1=z_1'=z_2=y_2=0,\;z_3=z_3'=z,\;y_1=z_2'=z'$ in the position space for the constitute quarks.
The propagator for the $\sigma$ meson exchange is given by
\begin{eqnarray}
\langle \sigma(z)\sigma(z')\rangle=\int \frac{d^4 p}{(2\pi)^4}\frac{i}{p^2-m^2_\sigma+im_\sigma\Gamma_\sigma}e^{-ip\cdot(z-z')}\,.
\end{eqnarray}
We  note that the factor of $(-1)$ in Eq. (\ref{Asigma}) comes from quark reordering,
$\Gamma_{\alpha\rho}=[\gamma_\mu(1-\gamma_5)]_{\alpha\rho}$, $\Gamma_{\beta\delta}=[\gamma^\mu(1-\gamma_5)]_{\beta\delta}$ from ${\cal H}_{\rm eff}$ in Eq. (\ref{Heff}), and $\Gamma^\sigma_{\gamma\gamma'}=\Gamma^\sigma_{\eta\eta'}=1$ from ${\cal L}_{\sigma qq}$ in Eq. (\ref{Lan}).
Note that the relevant Wilson coefficient is $(c_1-c_2)$ rather than $a_2=c_2+c_1/3$ due to the totally antisymmetric color indices in the baryon wave function and in the anti-triplet operator $(O_1-O_2)$, which is indeed the case found in the pole model calculation \cite{Cheng&Yang_PoleModel}.

To write down the matrix elements in Eq. (\ref{Asigma}) that are related to the wave functions of the $B$ meson and charmed baryons,  we first assign the four-momenta of $B^-$, $\Xi_c(2980)^0$, $\bar \Lambda_c$, and their constitute quarks as
\begin{eqnarray}\label{4momemtum}
&&B:\;\;\;\;\;\;\;\;\;\;p_B=(p_B^+,p_B^-,\vec{0}_\perp)\;,\;
\bigg\{
\begin{array}{l}
p_b=((1-\xi)p_B^+,(1-\xi)p_B^-,\vec{0}_\perp)\,,\\
p_l=(\xi p_B^+,\xi p_B^-,\vec{0}_\perp)\,,
\end{array}\nonumber\\
&&\Xi_c(2980):\;P=(p^+,p^-,\vec{0}_\perp)\;,\;
\Bigg\{
\begin{array}{l}
k_1=(x_1 p^+,p^-,\vec{k}_{1\perp})\,, \\
k_2=(x_2 p^+,0\;\;,\vec{k}_{2\perp})\,,  \\
k_3=(x_3 p^+,0\;\;,\vec{k}_{3\perp})\,,
\end{array}\nonumber\\
&&\bar\Lambda_c:\;\;\;\;\;\;\;\;\;\;P'=(p'^+,p'^-,\vec{0}_\perp)\;,\;
\Bigg\{
\begin{array}{l}
k'_1=(p'^+ ,x'_1p'^-,\vec{k'}_{1\perp})\,, \\
k'_2=(0\;\;,x'_2p'^-,\vec{k'}_{2\perp})\,,  \\
k'_3=(0\;\;,x'_3p'^-,\vec{k'}_{3\perp})\,,
\end{array}
\end{eqnarray}
where $x_i$ ($x'_i$) is the momentum fraction of the quark $i$ in the charmed baryon $\Xi_c(2980)$ ($\Lambda_c$), and $\vec{k'}_{i\perp}$ the corresponding transverse momenta. Note that the light-cone momenta
$p_B^{\pm}$ are equal to $m_B$  in the $B$ rest frame when the light quark masses are neglected.
As discussed in the Appendix, we will assume that $\Xi_c(2980)$ is a first positive-parity excitation with $J^P={1\over 2}$, $L_\ell=2$ and $J_\ell=1$, where $L_\ell$ and $J_\ell$ are the orbital and total angular momenta of the two light quarks of  $\Xi_c(2980)$.
In terms of the explicit four-momenta in Eq. (\ref{4momemtum}), the matrix elements involving $B^-$, $D$-wave $\Xi_c(2980)^0$ and $S$-wave $\Xi_c$ and $\overline{\Lambda}_c$ are given by
\begin{eqnarray}\label{Bwf1}
 \langle 0| \bar u_\eta^d(z') b^a_\rho(0)|B^-(p_B)\rangle
   &=&
   -i\frac{\delta^{da}}{3}\frac{f_B}{4}[(\not{\!p}_B+m_B)\gamma_5]_{\rho\eta}
   \int^1_0 d\xi e^{-i p_l\cdot z'}\Phi_B(\xi)\,,
 \nonumber\\
 \langle\Xi_c(2980)^0 (P)|\bar c_\alpha^a(0)\bar s_\beta^b(0)\bar d_\gamma^c(z)|0\rangle
   &=&
   \frac{\epsilon^{abc}}{6}\frac{f_{\Xi_c(2980)}}{4}
   [\bar u(P)\gamma_5\gamma_\mu]_\alpha \frac{1}{\sqrt {3}}\Bigg\{C^{-1}\Bigg[
   \sqrt{\frac{3}{20}}(\not{\!\tilde{k}}\tilde{K}^\mu+\not{\!\!\tilde{K}}\tilde{k}^\mu)
 \nonumber\\
  &-&\sqrt{\frac{2}{30}}\tilde{k}\cdot \tilde{K}(\gamma^\mu-\frac{P^\mu}{m_{\Xi'_c}})\Bigg]
   (\not{\!P}+m_{\Xi'_c})\Bigg\}_{\gamma\beta}\frac{2}{\beta^2}
 \nonumber \\
   &\times&\int [dx][d^2 k_\perp]e^{ik_3\cdot z}
   \Psi_{\Xi_c(2980)}(x_1,x_2,x_3,\vec{k}_{1\perp},\vec{k}_{2\perp},\vec{k}_{3\perp})\,,
 \nonumber\\
  \la \Xi^0_c(P)|\bar c^a_\alpha(0) \bar s_\beta^b(0) \bar d_\gamma^c(z)|0\ra
 &=&{\epsilon^{abc}\over 6}{f_{\Xi_c}\over4}
 \left[\bar u(P) \right]_\alpha
 \left[C^{-1}\gamma_5(P\!\!\!/+m_{\Xi_c})\right]_{\gamma\beta} \nonumber \\
 &\times&\int [dx][d^2 k_\perp]e^{ik_3\cdot z}
   \Psi_{\Xi_c}(x_1,x_2,x_3,\vec{k}_{1\perp},\vec{k}_{2\perp},\vec{k}_{3\perp})\,,
  \\
 \langle\overline{\Lambda}_c (P')|c_\delta^b(0) u_{\eta'}^d(z') d_{\gamma'}^c(z) |0\rangle
   &=&
   \frac{\epsilon^{bdc}}{6}\frac{f_{\Lambda_c}}{4}
   [\bar v(P')]_\delta[(\not{\!P}'-m_{\Lambda_c})\gamma_5 C]_{\eta'\gamma'}
   \nonumber \\
   &\times&\int [dx'][d^2 k'_\perp]e^{i(k'_2\cdot z'+k'_3\cdot z)}
   \Psi_{\Lambda_c}(x'_1,x'_2,x'_3,\vec{k}'_{1\perp},\vec{k}'_{2\perp},\vec{k}'_{3\perp})\,,
 \nonumber
\end{eqnarray}
with the decay constants $f_B$, $f_{\Xi_c(2980)}$,
$f_{\Lambda_c}$, the charge conjugation matrix $C$, and
\begin{eqnarray}
{[dx^{(\prime)}]}&=&dx_1^{(\prime)} dx_2^{(\prime)} dx_3^{(\prime)}\delta(1-\sum^{3}_{i=1}x_i^{(\prime)})\,,\nonumber\\
{[d^2 k_\perp^{(\prime)}]}&=&d^2 k_{1\perp}^{(\prime)}d^2 k_{2\perp}^{(\prime)}d^2 k_{3\perp}^{(\prime)}\delta^2(\vec{k}^{(\prime)}_{1\perp}+\vec{k}^{(\prime)}_{2\perp}+\vec{k}^{(\prime)}_{3\perp})
\,,\nonumber\\
  k&=&\frac{1}{2}(k_2-k_3),
  \quad
  K=\frac{1}{2}(k_2+k_3-2k_1)\,,
\end{eqnarray}
where $\tilde{A}\equiv A-P(P\cdot A)/m_{\Xi'_c}^2$ for $A=k$
or $K$. Recall that $k_2$ and $k_3$ are the 4-momenta of the two
light quarks in $\Xi_c(2980)$. The wave functions of $\Xi_c$ and $\Lambda_c$ are taken from \cite{wfswave}.
The derivation of the structures of the matrix elements involving the $D$-wave $\Xi_c(2980)$ is shown in the Appendix.

The amplitude $A_\sigma$ in Eq. (\ref{Asigma}) then becomes
\begin{eqnarray}
{\cal A}_\sigma &=&\frac{ig_\sigma^2}{18\times 4^3} \frac{G_F}{\sqrt 2}V_{cb}V_{cs}^*(c_1-c_2)f_B\;f_{\Xi_c(2980)}\;f_{\Lambda_c}\frac{2}{\sqrt {3}\beta^2}\int d\xi \int [dx][d^2 k_\perp][dx'][d^2 k'_\perp]\nonumber\\
&\times&(2\pi)^4\delta^4(k_3+k'_3+k'_2-p_l)\frac{1}{(k_3+k'_3)^2-m_\sigma^2+im_\sigma\Gamma_\sigma}\nonumber\\
&\times&\Phi_B(\xi)\Psi_{\Xi_c(2980)}(x_1,x_2,x_3,\vec{k}_{1\perp},\vec{k}_{2\perp},\vec{k}_{3\perp})
\Psi_{\Lambda_c}(x'_1,x'_2,x'_3,\vec{k}'_{1\perp},\vec{k}'_{2\perp},\vec{k}'_{3\perp})\nonumber\\
&\times&\bar u\gamma_5\gamma_\mu\Gamma[(\not{\!p}_B+m_B)\gamma_5]\Gamma^\sigma [(\not{\!\!P}'-m_{\Lambda_c})\gamma_5 C]\Gamma^\sigma\nonumber\\
&&\Bigg\{C^{-1}\Bigg[\sqrt{\frac{3}{20}}(\not{\!\tilde{k}}\tilde{K}^\mu+\not{\!\!\tilde{K}}\tilde{k}^\mu)-\sqrt{\frac{2}{30}}\tilde{k}\cdot \tilde{K}(\gamma^\mu-\frac{P^\mu}{m_{\Xi'_c}})\Bigg](\not{\!\!P}+m_{\Xi'_c})\Bigg\}\Gamma \,v\,,
\end{eqnarray}
where the delta function $\delta^4(k_3+k'_3+k'_2-p_l)$ in light-cone is presented as
\begin{eqnarray}
\delta^4(k_3+k'_3+k'_2-p_l)=-2\frac{1}{p^+}\frac{1}{p'^-}\delta(x_3 -\frac{\xi p_B^+}{p^+})\delta(x'_3 +x'_2 -\frac{\xi p_B^-}{p'^-} )\delta^2(\vec{k}'_{2\perp}+\vec{k}_{3\perp} +\vec{k}'_{3\perp} )\,.
\end{eqnarray}
After integrating over the variables with the $\delta$ functions,
we are led to
\begin{eqnarray}\label{Asigmaf}
{\cal A}_\sigma &=& \frac{ig_\sigma^2}{18\times 4^3} \frac{G_F}{\sqrt 2}V_{cb}V_{cs}^*(c_1-c_2)f_B\;f_{\Xi_c(2980)}\;f_{\Lambda_c}2(2\pi)^4\frac{1}{2^3}(2\pi)\int_0^{p'^-/p_B^-}d\xi \nonumber\\
&\times& \int_0^{1-\xi p_B^+/p^+} \frac{dx_2}{p^+} \int_0^{\xi p_B^-/p'^-} \frac{dx'_2}{p'^-}
\int^\infty_0 dk^2_{2\perp}\int^\infty_0 dk^2_{3\perp} \int^\infty_0 dk'^2_{3\perp}\int^{2\pi}_0 d \theta_{23}  \int^{2\pi}_0 d\theta_{33'} \nonumber\\
&\times&\Phi_B(\xi)\Psi_{\Xi_c(2980)}(x_1,x_2,x_3,\vec{k}_{1\perp},\vec{k}_{2\perp},\vec{k}_{3\perp})
\Psi_{\Lambda_c}(x'_1,x'_2,x'_3,\vec{k}'_{1\perp},\vec{k}'_{2\perp},\vec{k}'_{3\perp})\nonumber\\
&\times&\frac{\bar u(a_\sigma+b_\sigma\gamma_5)v}{(k_3+k'_3)^2-m_\sigma^2+im_\sigma\Gamma_\sigma}\,,
\end{eqnarray}
where
\begin{eqnarray}\label{ab1}
a_\sigma&=&\frac{12}{\sqrt
5\beta^2}\bigg[(k_{2\perp}^2-k_{3\perp}^2)+\frac{m_{\Xi'_c}^2}{4}(x_2^2-x_3^2)\bigg]m_{\Xi'_c}(m_B+m_{\Lambda_c}+m_{\Xi'_c})
(m_B+m_{\Lambda_c}-m_{\Xi'_c}) \,,\nonumber\\
b_\sigma&=&\frac{-12}{\sqrt
5\beta^2}\bigg[(k_{2\perp}^2-k_{3\perp}^2)+\frac{m_{\Xi'_c}^2}{4}(x_2^2-x_3^2)\bigg]m_{\Xi'_c}^2
m_{\Lambda_c}\,,
\end{eqnarray}
with
\begin{eqnarray}
&&x_1^{(\prime)}=1-x_2^{(\prime)}-x_3^{(\prime)},\;x_3=\frac{\xi p_B^+}{p^+},\;x_3'=\frac{\xi p_B^-}{p'^-}-x_2',\nonumber\\
&&\vec{k}_{1\perp}=-(\vec{k}_{2\perp}+\vec{k}_{3\perp}),\;\vec{k}_{1\perp}^{\prime}=\vec{k}_{3\perp}.
\end{eqnarray}
Note that $\theta_{23}$ and $\theta_{33'}$ are the angles of
$\vec{k}_{2\perp}$ and of $\vec{k}'_{3\perp}$ as measured against
$\vec{k}_{3\perp}$, respectively. We can also obtain $A_{\pi^0}$
and $A_{\pi^\pm}$ in Eq. (\ref{amptotal}) by replacing the
notation of $\sigma$ in Eq. (\ref{Asigmaf}) by $\pi^0$ and
$\pi^\pm$, respectively, $a_\sigma$ and $b_\sigma$ by
\begin{eqnarray}\label{ab2}
a_{\pi^{0(\pm)}}&=&\frac{-12}{\sqrt
5\beta^2}\bigg[(k_{2\perp}^2-k_{3\perp}^2)+\frac{m_{\Xi'_c}^2}{4}(x_2^2-x_3^2)\bigg]m_{\Xi'_c}
(m_B-m_{\Lambda_c}+m_{\Xi'_c})(m_B+m_{\Lambda_c}-m_{\Xi'_c}) \,,\nonumber\\
b_{\pi^{0(\pm)}}&=&\frac{12}{\sqrt
5\beta^2}\bigg[(k_{2\perp}^2-k_{3\perp}^2)+\frac{m_{\Xi'_c}^2}{4}(x_2^2-x_3^2)\bigg]m_{\Xi'_c}(m_B^2+
m_{\Lambda_c}^2-m_{\Xi'_c}^2+m_{\Xi'_c} m_{\Lambda_c})\,,
\end{eqnarray}
and
$\Gamma_{\eta\eta'}^{\pi^0}=-\Gamma_{\eta\eta'}^{\pi^\pm}=-\Gamma_{\gamma\gamma'}^{\pi^0}
=\Gamma_{\gamma\gamma'}^{\pi^\pm}=i\gamma_5$. The amplitude of
$\bar B^0\to \Lambda_c^+\bar \Lambda_c^-$ is similar to that of
$\bar B\to \Xi_c \bar \Lambda_c$ studied in \cite{ChengChuaTsi}
except for the CKM matrix element being replaced by
$V_{cb}V^*_{cd}$ , and its $a_\sigma$, $b_\sigma$,
$a_{\pi^{0(\pm)}}$, $b_{\pi^{0(\pm)}}$  are given by
\begin{eqnarray}
a_\sigma&=&-4m_{\Lambda_c}m_B(m_B+2m_{\Lambda_c})\,,\nonumber\\
b_\sigma&=&\;\;\;4m_{\Lambda_c}(m_B-2m_{\Lambda_c})(m_B+2m_{\Lambda_c})\,,\nonumber\\
a_{\pi^{0(\pm)}}&=&\;\;\;4m_{\Lambda_c}m_B^2\,,\nonumber\\
b_{\pi^{0(\pm)}}&=&-4m_{\Lambda_c}m_B(m_B-2m_{\Lambda_c})\,.
\end{eqnarray}
Once the explicit expressions for the wavefunctions $\Phi_B$, $\Psi_{\Xi_c(2980)}$,
$\Psi_{\Lambda_c}$ and other parameters are given, we are ready
to carry out the numerical analysis.

\section{numerical analysis}

To proceed with the numerical calculations, we need to specify the
relevant wave functions. For the $B$ meson, it is given by
\cite{wavefnB}
\begin{eqnarray}
\Phi_B(\xi)=N_B \xi^2(1-\xi^2)\text{exp}\bigg[-\frac{1}{2}\frac{\xi^2 m_B^2}{\omega_B^2}\bigg]\,,
\end{eqnarray}
with $\omega_B=0.38\pm 0.04$ GeV, where $N_B$ is determined by the normalization
\begin{eqnarray}
\int^1_0 d\xi \Phi_B(\xi)=1\,.
\end{eqnarray}
For the charmed baryon, such as $D$-wave $\Xi_c(2980)$ and
$S$-wave $\Lambda_c$, we assume that their wave functions have
similar expression \cite{wavefnBcBc}
\begin{eqnarray}
\Psi_{\bf B_c}(x_i,\vec{k}_{i\perp})
=\frac{N_{\bf B_c}}{(2\pi \beta^2)^2}
\prod^3_{i=1}\text{exp}\bigg[-\frac{\vec{k}_{i\perp}^2+\hat{m}_i^2}{2\beta^2 x_i}\bigg]\,,
\end{eqnarray}
with $\beta=0.96\pm 0.04$ GeV and $\hat{m}_i$ the mass of the constitute
quark $i$, where $N_{\bf B_c}$ is given by the normalization
\begin{eqnarray}
\int[dx][dk_{\perp}^2]\Psi_{\bf B_c}(x_i,\vec{k}_{i\perp})
=\int[dx]N_{\bf B_c}\prod^{3}_{i=1} x_i\;\text{exp}\bigg[-\frac{\hat{m}_i^2}{2\beta^2 x_i}\bigg]=1\,.
\end{eqnarray}
For the decay constants, $f_{\Lambda_c}$ can be related to the
decay constant of the $\Lambda_b$  by the relation $f_{\bf B_c}
m_{\bf B_c}=f_{\Lambda_b} m_{\Lambda_b}$ \cite{Bcdecay}, and we
let $f_{\Xi_c(2980)}\simeq f_{\Xi_c}$ due to the lack of
information on the decay constant of the $D$-wave charmed baryon.
For other input parameters, see Table \ref{input}.

\begin{table}[h!]
\caption{Summary of the input parameters.}\label{input}
\begin{tabular}{|l|l|}
\hline
$\omega_B=0.38\pm 0.04$ GeV                       &$f_B=0.2$ GeV  \\
$\beta=0.96\pm 0.04$    GeV                       &$f_{\Xi_c(2980)}\simeq f_{\Xi_c}$\\
$\hat{m}_s=0.46\pm 0.06$      GeV                 &$f_{\Xi_c}=6.2\times 10^{-3}$ GeV$^2$\\
$\hat{m}_{u(d)}=0.26\pm 0.04$ GeV                 &$f_{\Lambda_c}=6.7\times 10^{-3}$ GeV$^2$\\
$m_{\Xi_c(2980)}=2.93$ GeV \cite{LambdaCLambdaCK} &$\Gamma_\sigma=0.6$ GeV\\
$g_\sigma$=3.35 \cite{ChengChuaTsi}               &$\Gamma_{\pi^0}=7.8\times 10^{-9}$ GeV\\
$g_\pi=\sqrt 2 g_{\pi^0}$=4.19 \cite{ChengChuaTsi}&$\Gamma_{\pi^\pm}=2.5\times 10^{-17}$ GeV\\\hline
\end{tabular}
\end{table}

For the two-body baryonic B decay
amplitude given by
\begin{eqnarray}
A(B\to {\bf B_c \bar B_c'})=\bar u(A+B\gamma_5)v\,,
\end{eqnarray}
the decay rate reads  \cite{Dstopn}
\begin{eqnarray}\label{Gamma}
\Gamma(B\to {\bf B_c \bar B_c'})=\frac{p_c}{4\pi m_B^2}\bigg\{
|A|^2[m_B^2-(m_{\bf B_c'}+m_{\bf B_c})^2]+|B|^2[m_B^2-(m_{\bf B_c'}-m_{\bf B_c})^2]\bigg\}\,,
\end{eqnarray}
where $p_c$ is the c.m. momentum.  To obtain the rate for $\bar B\to \Lambda_c^+\bar \Lambda_c^-
\bar K$, we shall use $\B(\Xi_c(2980)\to \Lambda_c \bar K)=0.5$ derived from the $^3P_0$ model
\cite{decaywidth}. The calculated results for $\bar B\to \Xi_c\bar\Lambda_c$, $\bar B^0\to
\Lambda_c^+\bar \Lambda_c^-$ and $\bar B\to \Lambda_c^+\bar
\Lambda_c^- \bar K$ are summarized in Table \ref{tabC}.

\begin{table}[h!]
\caption{Branching ratios (in units of $10^{-4}$) of doubly charmful two-body and three-body baryonic $B$ decays, where the first and second theoretical errors come from $\beta$ and $\omega_B$,
while the third and fourth errors are from $\hat{m}_{u(d)}$ and $\hat{m}_s$, respectively. Use of $\B(\Xi_c(2980)\to\Lambda_c\bar K)=0.5$ has been made to derive the rate of $\bar B\to\Lambda_c\bar\Lambda_c\bar K$. }\label{tabC}
\begin{tabular}{|l|c||c|c|c|}
\hline
  &Theory&BaBar&Belle&Average\\\hline
$B^-\to \Xi_c^0\bar \Lambda_c^-$                     &$10.4^{+3.8}_{-3.6}$$^{+0.3}_{-1.8}$$^{+4.2}_{-3.5}$$^{+0.3}_{-1.3}$
&$16\pm 7^{+9}_{-4}$&$37\pm 15^{+21}_{-\;\,9}$&~~$20\pm 6^{+11}_{-\;\,5}$~~\\
$\bar B^0\to \Xi_c^+\bar \Lambda_c^-$                &\;\;$9.4^{+4.6}_{-2.6}$$^{+0.4}_{-0.8}$$^{+4.3}_{-3.0}$$^{+0.5}_{-0.4}$
&~~$3.8\pm 3.1^{+8.7}_{-2.4}<14$~~&$24\pm 12^{+53}_{-15}$&---\\
$B^-\to \Lambda_c^+\bar \Lambda_c^- K^-$             &\;\;$3.6^{+1.0}_{-1.0}$$^{+0.8}_{-1.0}$$^{+1.5}_{-1.2}$$^{+0.5}_{-0.7}$
&$11.4\pm 6.4$ &$6.5\pm 3.7$                &$7.7\pm 3.2$\\
$\bar B^0\to \Lambda_c^+\bar \Lambda_c^- \bar K^0$   &\;\;$3.3^{+1.2}_{-0.9}$$^{+0.8}_{-0.9}$$^{+0.9}_{-1.1}$$^{+0.2}_{-0.6}$
&$3.8\pm 3.0$  &$7.9\pm 5.2$                &$5.2\pm 3.0$ \\
$\bar B^0\to \Lambda_c^+\bar \Lambda_c^-$            & ~~$0.52^{+0.23}_{-0.11}$$^{+0.06}_{-0.03}$$^{+0.26}_{-0.15}$$^{+0}_{-0}$~~
&---              &~~$0.22^{+0.26}_{-0.21}<0.62$~~ & $<0.62$ \\\hline
\end{tabular}
\end{table}

\section{discussion and conclusion}

In this work we have studied the doubly charmful two-body and
three-body baryonic $B$ decays $B\to \Lambda_c^+\bar \Lambda_c^-$
and $\bar B\to \Lambda_c^+\bar \Lambda_c^-\bar K$. For the former decay, our prediction for its branching ratio of order $5\times 10^{-5}$ (see Table \ref{tabC}) is consistent with the
extrapolation from $\B(\bar B^0\to
\Xi_c^+\bar \Lambda_c^-)$ and from $\B(\bar B^0\to \Lambda_c\bar
p)$ provided that the dynamical suppression of $\Lambda_c\bar p$ relative to $\Lambda_c\bar\Lambda_c$ is taken into account.
As pointed out before, a
naive estimate of the branching ratio ${\cal O}(10^{-8})$ for the decay $\bar B\to \Lambda_c^+\bar \Lambda_c^-\bar K$ is too small by three to four orders of magnitude
compared to experiment. Previously, it has been shown that a large
enhancement for the $\Lambda_c^+\bar\Lambda_c^-\bar K$ production can
occur due to a charmonium-like resonance (for example, the $X(4630)$ state
discovered by Belle) with a mass near the $\Lambda_c\bar\Lambda_c$
threshold. Motivated by the BaBar's observation of a resonance in
the $\Lambda_c\bar K$ system with a mass of order 2930 MeV,  we have studied
the contribution to $B\to\Lambda_c^+\bar\Lambda_c^-K$
from the intermediate state $\Xi_c(2980)$  which is postulated to
be a first positive-parity excited $D$-wave charmed baryon state.  Assuming
that a soft $q\bar q$ quark pair is produced through the $\sigma$
and $\pi$ meson exchanges in the configuration for $B\to
\Xi_c(2980)\bar\Lambda_c$, it is
found that the branching ratio of  $\bar B\to \Lambda_c^+\bar \Lambda_c^-\bar
K$ is of order $3.5\times 10^{-4}$. This is  in agreement with experiment for $\Lambda_c\bar\Lambda_c\bar K^0$, but slightly smaller for $\Lambda_c\bar\Lambda_c K^-$. In conjunction with the previous analysis \cite{ChengChuaTsi}, we conclude that the enormously large rate of $\bar B\to \Lambda_c^+\bar \Lambda_c^-\bar K$ arises from the resonances $\Xi_c(2980)$ and $X(4630)$.

We have also presented updated results for $\bar B\to \Xi_c\bar\Lambda_c$ which are slightly smaller than our previous analysis \cite{ChengChuaTsi} but are consistent with experiment.

\section*{Acknowledgments}

This work is supported in part by the National Science Council of
R.O.C. under Grants No. NSC97-2112-M-001-004-MY3 and
NSC97-2112-M-033-002-MY3.

\appendix

\section{\label{sec:dwave}}

It has been conjectured that $\Xi_c(2980)$ is likely to be a first
positive-parity excitation with $J^P={1\over 2}^+$ \cite{Dwave}.
Denoting the quantum numbers $L_k$ and $L_K$ as the eigenvalues of
$\vec{L}_k^2$ and $\vec{L}_K^2$, respectively, the $k$-orbital
momentum $L_k$ describes relative orbital excitations of the two
light quarks, and the $K$-orbital momentum $L_K$ describes orbital
excitations of the center of the mass of the two light quarks
relative to the heavy quark \cite{Korner94}. The first positive-parity 
excitations are those states with $L_K+L_k=2$. According to
Table IV of \cite{Dwave}, possible antitriplet candidates for
$\Xi_c(2980)$ are $\tilde\Xi_{c1}({1\over
2}^+),\tilde\Xi''_{c0}({1\over 2}^+),\tilde\Xi''_{c1}({1\over
2}^+)$ and $\tilde\Xi'''_{c1}({1\over 2}^+)$, where the quantum
number in the subscript labels $J_\ell$, the total angular
momentum of the two light quarks. (We use a tilde to denote states
with antisymmetric orbital wave functions (i.e. $L_K=L_k=1$) under
the interchange of two light quarks.) Strong decays of these four
states have been studied in \cite{decaywidth} using the $^3P_0$
model. It turns out that $\Gamma(\tilde\Xi_{c1}({1\over
2}^+))\approx 3.2$ MeV is too small and
$\Gamma(\tilde\Xi''_{c1}({1\over 2}^+))\approx 148$ MeV is too
large compared to the experimental value of order 30 MeV [see Eq. (\ref{width})], while
$\tilde\Xi''_{c0}({1\over 2}^+)$ does not decay into $\Xi_c\pi$
and $\Lambda_c\bar K$. Therefore, the favored candidate is
$\tilde\Xi'''_{c1}({1\over 2}^+)$ which has $L_\ell=2$ and
$J_\ell=1$.

We use the light-front approach to obtain the structure of the
matrix element for the $D$-wave charmed baryon as shown in
Eq.~(\ref{Bwf1}). In the light-front formalism, the charmed baryon
bound state with the total momentum $P$, spin $J=1/2$ and the
angular momentum of the light quark pair $\vec J_l\equiv \vec
S_l+\vec L_l$ with $\vec S_l\equiv \vec S_2+\vec S_3$ and $\vec
L_l\equiv \vec L_k+\vec L_K$ can be written as (see, for example
\cite{Cheng97,CCH,pentaquark})
\begin{eqnarray}
        |{\bf B}_c(P,\{L_k,L_K,L_l,S_l,J_l\},J,,J_z)\rangle
                &=&\int \{d^3p_1\}\{d^3p_2\}\{d^3p_3\} ~\frac{2(2\pi)^3}{\sqrt {P^+}}
                \delta^3(\bar P -\bar p_1-\bar p_2-\bar p_3)
                ~\nonumber\\
        &\times& \sum_{\lambda_i,\alpha,\beta,\gamma,a,b}
                \Psi^{J J_z}_{\{L\}}(x_1,x_2,x_3,k_{1\bot},k_{2\bot},k_{3\bot},\lambda_1,\lambda_2,\lambda_3)~
        \non\\
        &\times& ~ C^{abc} F_{ff'}^{\{L\}}
             \Big|c^a(p_1,\lambda_1) q_f^b(p_2,\lambda_2) q_{f'}^c(p_3,\lambda_3)\Big\rangle,
 \label{lfmbs}
\end{eqnarray}
where $a,b,c$ and $f,f'$ are color and flavor indices,
respectively, $\lambda_{1,2,3}$ denote helicities, $\bar p_1$,
$\bar p_2$ and $\bar p_3$ are the on-mass-shell light-front
momenta,
\begin{equation}
        \bar p=(p^+, p_\bot)~, \quad p_\bot = (p^1, p^2)~,
                \quad p^- = {m^2+p_\bot^2\over p^+},
\end{equation}
and
\begin{eqnarray}
        &&\{d^3p\} \equiv {dp^+d^2p_\bot\over 2(2\pi)^3},
        \quad \delta^3(\bar p)=\delta(p^+)\delta^2(p_\bot),
        \nonumber \\
        &&\Big|c(p_1,\lambda_1) q_f(p_2,\lambda_2) q_{f'}(p_3,\lambda_3)\Big\rangle
        = b^\dagger_{c\lambda_1}(p_1) b^\dagger_{q_f\lambda_2}(p_2) b^\dagger_{q_{f'}\lambda_3}(p_3)|0\rangle,\\
        &&\{b_{q'\lambda'}(p'),b_{q\lambda}^\dagger(p)\} =
        2(2\pi)^3~\delta^3(\bar p'-\bar p)~\delta_{\lambda'\lambda}~\delta_{q'q}.
                \nonumber
\end{eqnarray}
The coefficients $C^{abc}=\epsilon^{abc}/6$ and $F_{ff'}^{\{L\}}$
are normalized color factor and flavor coefficient, respectively.

In terms of the light-front relative momentum variables $(x_i,
k_{i\bot})$ for $i=1,2,3$ defined by
\begin{eqnarray}
        && p^+_i=x_i P^{+}, \quad \sum_{i=1}^3 x_i=1, \nonumber \\
        && p_{i\bot}=x_i P_\bot+k_{i\bot}, \quad \sum_{i=1}^3 k_{i\bot}=0,
\end{eqnarray}
the momentum-space wave-function $\Psi^{JJ_z}_{\{L\}}$ can be
expressed as
\begin{eqnarray}
        \Psi^{JJ_z}_{\{L\}}(x_i,k_{i\bot},\lambda_i)
                &=& \left(\Pi_{i=1}^3 \langle \lambda_i|{\cal R}_M^\dagger(x_i,k_{i\bot}, m_i)|s_i\rangle\right)
                \la J_l S_1; j_l s_1|J_l S_1;J J_z\ra
                \non\\
                &\times&
                       \la S_l L_l; s_l l_l|S_l L_1;J_l j_l\ra
                       \la L_k L_K; l_k l_K|L_k L_K;L_l l_l\ra
                       \la S_2 S_3; s_2 s_3|S_2 S_3;S_l s_l\ra
                       \non\\
                  &\times& \phi_{L_k L_K,l_k l_K}(x_1,x_2,x_3,k_{1\bot},k_{2\bot},k_{3\bot}),
\label{eq:Psi}
\end{eqnarray}
where $\phi_{l_k l_K}(x_1,x_2,x_3,k_{1\bot},k_{2\bot},k_{3\bot})$
describes the momentum distribution of the constituents in the
bound state with the subsystem consisting of the particles 2 and 3
in the orbital angular momentum $(L_k)_z=l_k,\,(L_K)_z=l_K$ state,
$\la J_l S_1; j_l s_1|J_l S_1;J J_z\ra$, and so on are
Clebsch-Gordan coefficients and $\langle \lambda_i|{\cal
R}_M^\dagger(x_i,k_{1\bot}, m_i)|s_i\rangle$ is the well
normalized Melosh transform matrix element. Explicitly
\cite{Jaus90,deAraujo:1999cr},
 \be
        \la \lambda_i|{\cal R}^\dagger_M (x_i,k_{i\bot},m_i)|s_i\ra
        &=&\frac{\bar
        u(k_i,\lambda) u_D(k_i,s_i)}{2 m_i}
        =-\frac{\bar
        v(k_i,\lambda) v_D(k_i,s_i)}{2 m_i}
        \non\\
        &=&\frac{(m_i+x_i M_0)\delta_{\lambda_i s_i}
                      -i\vec \sigma_{\lambda_i s_i}\cdot\vec k_{i\bot} \times
                      \vec                n}
                {\sqrt{(m_i+x_i M_0)^2 + k^{2}_{i\bot}}},
 \en
with $u_{(D)}$ and $v_{(D)}$ Dirac spinors in the light-front
(instant) form, $\vec n = (0,0,1)$, a unit vector in the
$z$-direction, and
 \be
 M_0^2&=&\sum_{i=1}^3\frac{m_i^2+k^2_{i\bot}}{x_i},\quad
 k_i=(\frac{m_i^2+k^2_{i\bot}}{x_i M_0},x_i M_0,\,
 k_{i\bot})=(e_i-k_{iz},e_i+k_{iz},k_{i\bot}),
 \non\\
 M_0&=&e_1+e_2+e_3,\quad
 e_i =\sqrt{m^{2}_i+k^{2}_{i\bot}+k^{2}_{iz}}=\frac{x_i M_0}{2}+\frac{m_i^2+k^{2}_{i\bot}}{2 x_i M_0},\quad
 k_{iz}=\frac{x_i M_0}{2}-\frac{m_i^2+k^{2}_{i\bot}}{2 x_i M_0}.
 \non\\
 \en
Note that
 $u_D(k_i,s_i)=u(k_i,\lambda_i) \la \lambda_i|{\cal R}^\dagger_M|s_i\ra$
and, consequently, the state $|q(k_i,\lambda_i)\ra \la
\lambda_i|{\cal R}^\dagger_M|s_i\ra$ transforms like
$|q(k_i,s_i)\ra$ under rotation, i.e. its transformation does not
depend on its momentum. A crucial feature of the light-front
formulation of a bound state, such as the one shown in
Eq.~(\ref{lfmbs}), is the frame-independence of the light-front
wave function~\cite{Brodsky:1997de,Jaus90}. Namely, the hadron can
be boosted to any (physical) ($P^+$, $P_\bot$) without affecting
the internal variables ($x_i$, $k_{\bot i}$) of the wave function,
which is certainly not the case in the instant-form formulation.

In practice it is more convenient to use the covariant form for
the Melosh transform matrix element
\begin{eqnarray}
       &&\langle \lambda_1|{\cal R}_M^\dagger(x_1,k_{1\bot}, m_1)|s_1\rangle
                \la J_l S_1; j_l s_1|J_l S_1;J J_z\ra
             =\frac{M_0}{\sqrt{2(p_1\cdot\bar P+m_1 M_0)}}
        ~\bar u(p_1,\lambda_1)\Gamma_{J_lj_l} u_{\bf B_c}(\bar P,J_z),
        \non\\
        &&\langle \lambda_2|{\cal R}_M^\dagger(x_2,k_{2\bot}, m_2)|s_2\rangle
        \langle \lambda_3|{\cal R}_M^\dagger(x_3,k_{3\bot}, m_3)|s_3\rangle
                \la S_2 S_3; s_2 s_3|S_2 S_3;S_l s_l\ra
       \non\\
       &&\qquad=\frac{1}{2 M_0\sqrt{2(p_2\cdot\bar P+m_2 M_0)(p_2\cdot\bar P+m_2 M_0)}}
        ~\bar u(p_2,\lambda_2)(\not\!\bar P+M_0)\gamma_5\Gamma_{S_ls_l} v(p_3,\lambda_3),
        \label{eq:covariant}
\end{eqnarray}
with $S_i=J=1/2$ and
\begin{eqnarray}
        &&\Gamma_{00}=1,\qquad
        \Gamma_{1m}=-\frac{1}{\sqrt3}\gamma_5\not\!\varepsilon^*(\bar
        P,m),\non\\
        &&\bar P\equiv \bar p_1+\bar p_2+\bar p_3, \non \\
        &&\varepsilon^\mu(\bar P,\pm 1) =
                \left[{2\over P^+} \vec \varepsilon_\bot (\pm 1) \cdot
                \vec P_\bot,\,0,\,\vec \varepsilon_\bot (\pm 1)\right],
                \quad \vec \varepsilon_\bot
                (\pm 1)=\mp(1,\pm i)/\sqrt{2}, \nonumber\\
        &&\varepsilon^\mu(\bar P,0)={1\over M_0}\left({-M_0^2+P_\bot^2\over
                P^+},P^+,P_\bot\right),   \label{polcom}
\end{eqnarray}
for states with $J_l,S_l=0$ or $1$,  see \cite{CCH,pentaquark} for
the derivation the above expressions.
The following identities will be useful later:
 \be
 &&\la 11;m' m''|11;2m\ra
 \la 12;m''' m|12;1 n\ra
 ~\varepsilon_\mu(\bar P,m')\varepsilon_\nu(\bar P,m'')\varepsilon_\rho(\bar P,m''')
 \non\\
 &&\qquad\qquad\qquad\qquad=-\sqrt{\frac{3}{20}}
    \left[\varepsilon^*_\mu(\bar P,m)\varepsilon_\nu(\bar P,n)\varepsilon_\rho(\bar P,m)+
     \varepsilon_\mu(\bar P,n)\varepsilon^*_\nu(\bar P,m)\varepsilon_\rho(\bar P,m)\right]
  \non\\
 &&\qquad\qquad\qquad\qquad\quad  +\sqrt{\frac{2}{30}}\varepsilon^*_\mu(\bar P,m)\varepsilon_\nu(\bar P,m)\varepsilon_\rho(\bar
 P,n),\non\\
 &&\varepsilon^*_\mu(\bar P,m)\varepsilon_\nu(\bar P,m)
 =-g_{\mu\nu}+\frac{\bar P_\mu\bar P_\nu}{M_0^2},
 \en
where $\epsilon^*(m)=(-)^m\epsilon(-m)$ is used in the first
identity, and $\tilde A_\mu$ is defined as
$-\varepsilon^*_\mu(\bar P,m)[\varepsilon(\bar P,m)\cdot A]$ for
later purposes.

Under the constraint of $1-\sum_{i=1}^3 x_i=\sum_{i=1}^3
(k_i)_{x,y,z}=0$, we have the expressions
 \be
   &&\phi_{L_kL_K,l_kl_K}(\{x\},\{k_\bot\})
   =\left(\frac{3}{2}\right)^{3/2}\sqrt{\frac{\partial(k_{2z}, k_{3z})}{\partial
   (x_2,
  x_3)}}\,\varphi_{L_K l_K}(\vec K,\beta_K)~\varphi_{L_k l_k}(\vec k,\beta_k), \non\\
  &&
  \qquad\qquad\varphi_{00}(\vec \kappa,\beta)=\varphi(\vec \kappa,\beta),\quad
  \varphi_{1m}(\vec \kappa,\beta)=\kappa_{m} \varphi_p(\vec \kappa,\beta),
  \label{eq:phi}
 \en
where $\kappa_m=\vec\varepsilon(\bar P, m)\cdot\vec \kappa$, or
explicitly $\kappa_{m=\pm1}=\mp(\kappa_{\bot x}\pm i \kappa_{\bot
y})/\sqrt2$, $\kappa_{m=0}=\kappa_{z}$ with $\kappa_{x,y,z}$ in
the rest frame of $\bar P$, are proportional to the spherical
harmonics $Y_{1m}$ in the momentum space, and $\varphi$,
$\varphi_p$ are the distribution amplitudes of $S$-wave and
$P$-wave states, respectively, and the factor $(3/2)^{3/2}
\sqrt{\partial(k_{2z}, k_{3z})/\partial (x_2, x_3)}$ in
Eq.~(\ref{eq:phi}) is a normalization factor. For a Gaussian-like
wave function, one has \cite{Cheng97,CCH}
\begin{eqnarray} \label{eq:Gauss}
 \varphi(\vec \kappa,\beta)
    &=&4 \left({\pi\over{\beta^{2}}}\right)^{3\over{4}}
               ~{\rm exp}
               \left(-{\kappa^2_z+\kappa^2_\bot\over{2
               \beta^2}}\right),\quad
    \varphi_p(\vec \kappa,\beta)=\sqrt{2\over{\beta^2}}~\varphi(\vec \kappa,\beta).
 \label{eq:wavefn}
\end{eqnarray}

By using the above equations it is straightforward to obtain
 \be
 &&(\Pi_i\langle \lambda_i|{\cal R}_M^\dagger(x_i,k_{i\bot}, m_i)|s_i\rangle)~
 \la J_l S_1; j_l s_1|J_l S_1;J J_z\ra
                \la S_l L_l; s_l l_l|S_l L_1;J_l j_l\ra
 \non\\
       &&\qquad\times      \la L_k L_K; l_k l_K|L_k L_K;L_l l_l\ra
               \la S_2 S_3; s_2 s_3|S_2 S_3;S_l s_l\ra
               \\
 &&\qquad\qquad=\frac{1}{4\sqrt{\Pi_i(p_i\cdot\bar P+m_i M_0)}}
        ~\bar u(p_1,\lambda_1)u_{\bf B_c}(\bar P,J_z)
        ~\bar u(p_2,\lambda_2)(\not\!\bar P+M_0)\gamma_5 C \bar
        u^T(p_3, \lambda_3) \non
         \label{eq:S}
 \en
for $J_l=S_l=L_l=L_k=L_K=0$ and
 \be
 &&(\Pi_i\langle \lambda_i|{\cal R}_M^\dagger(x_i,k_{i\bot}, m_i)|s_i\rangle)~
 \la J_l S_1; j_l s_1|J_l S_1;J J_z\ra
                \la S_l L_l; s_l l_l|S_l L_1;J_l j_l\ra
 \non\\
       &&\qquad\times      \la L_k L_K; l_k l_K|L_k L_K;L_l l_l\ra
               \la S_2 S_3; s_2 s_3|S_2 S_3;S_l s_l\ra
               \frac{\sqrt2 k_{l_k}}{\beta_k}
               \frac{\sqrt2 K_{l_K}}{\beta_K}
               \non\\
 &&\qquad\qquad=\frac{-1}{\beta_k\beta_K\sqrt{12\,\Pi_i(p_i\cdot\bar P+m_i M_0)}}
        ~\bar u(p_1,\lambda_1)\gamma_5\gamma_\mu u_{\bf B_c}(\bar P,J_z)
 \\
 &&\qquad\qquad     \times~ \bar u(p_2,\lambda_2)(\not\!\bar P+M_0)
 \left\{\sqrt{\frac{3}{20}}(\not{\!\tilde{k}}\tilde{K}^\mu+\not{\!\!\tilde{K}}\tilde{k}^\mu)
   -\sqrt{\frac{2}{30}}\tilde{k}\cdot \tilde{K}(\gamma^\mu-\frac{\bar P^\mu}{M_0})\right\} C \bar
        u^T(p_3, \lambda_3)   \non
         \label{eq:D}
 \en
for $L_l=2$, $L_k=L_K=S_l=1$. Note that the factors of
$k_m=\varepsilon(\bar P,m)\cdot(\bar p_2-\bar p_3)/2$, and
$K_m=\varepsilon(\bar P,m)\cdot(\bar p_2+\bar p_3-2\bar p_1)/2$
come from the wave function Eq.~(\ref{eq:phi}) for the $L_k=L_K=1$
case. Promoting $\bar P\to P$ and $M_0\to M$ and taking Hermitian
conjugation (to change the initial state to the final state)
we obtain the structure of the matrix element for the $S$-wave and
$D$-wave charmed baryons as shown in Eq.~(\ref{Bwf1}). Note that
for simplicity we have taken $\beta=\beta_k=\beta_K$ in
Eq.~(\ref{Bwf1}).


\begin{thebibliography}{99}

\bibitem{Suzuki}
M.~Suzuki, J.\ Phys.\ G {\bf 34}, 283 (2007).

\bibitem{exampleC}
S.~A.~Dytman {\it et al.}  [CLEO Collaboration], Phys.\ Rev.\  D {\bf 66}, 091101 (2002);
N.~Gabyshev {\it et al.}  [Belle Collaboration], Phys.\ Rev.\  D {\bf 66}, 091102 (2002);
N.~Gabyshev {\it et al.}  [Belle Collaboration], Phys.\ Rev.\ Lett.\  {\bf 97}, 242001 (2006);
K.~S.~Park {\it et al.}  [Belle Collaboration], Phys.\ Rev.\  D {\bf 75}, 011101 (2007).

\bibitem{LambdaCbarp}
P.~Krokovny {\it et al.}  [Belle Collaboration], Phys.\ Rev.\ Lett.\  {\bf 90}, 141802 (2003);
B.~Aubert {\it et al.}  [BaBar Collaboration], Phys.\ Rev.\  D {\bf 78}, 112003 (2008).

\bibitem{ppD}
K.~Abe {\it et al.} [Belle Collaboration], Phys. Rev. Lett. {\bf 89}, 151802 (2002);
B.~Aubert {\it et al.}  [BaBar Collaboration], Phys.\ Rev.\  D {\bf 74}, 051101 (2006).

\bibitem{LambdaCLambdaCK}
B.~Aubert {\it et al.}  [BaBar Collaboration], Phys.\ Rev.\  D {\bf 77}, 031101 (2008).

\bibitem{LambdaCLambdaC}
K.~Abe {\it et al.}  [Belle Collaboration], Phys.\ Rev.\  D {\bf 77}, 051101 (2008).

\bibitem{LambdaCLambdaCK0}
K.~Abe {\it et al.}  [Belle Collaboration], Phys.\ Rev.\ Lett.\  {\bf 97}, 202003 (2006).

\bibitem{ChengChuaTsi}
H.~Y.~Cheng, C.~K.~Chua and S.~Y.~Tsai, Phys.\ Rev.\  D {\bf 73}, 074015 (2006).

\bibitem{Cheng&Tseng}
H.~Y.~Cheng and B.~Tseng, Phys.\ Rev.\  D {\bf 48}, 4188 (1993).

\bibitem{Korner} J.G. K\"orner and M. Kr\"amer, Z. Phys. C {\bf 55},
659 (1992); Q.P. Xu and A.N. Kamal, Phys. Rev. D {\bf 46}, 270
(1992); P. \.Zenczykowski, Phys. Rev. D {\bf 50}, 402
(1994).

\bibitem{Hou&Soni}
W.~S.~Hou and A.~Soni, Phys.\ Rev.\ Lett.\  {\bf 86}, 4247 (2001).

\bibitem{He}
X.~G.~He, T.~Li, X.~Q.~Li and Y.~M.~Wang, Phys.\ Rev.\  D {\bf 75}, 034011 (2007).

\bibitem{BelleLambdac} G.~Pakhlova {\it et al.}  [Belle Collaboration],
  Phys.\ Rev.\ Lett.\  {\bf 101}, 172001 (2008).

\bibitem{BelleY4660} X.L. Wang {\it et al.}  [Belle Collaboration],
  Phys.\ Rev.\ Lett.\  {\bf 99}, 142002 (2007).

\bibitem{Chen}
C.~H.~Chen, Phys.\ Lett.\  B {\bf 638}, 214 (2006).

\bibitem{pdg}
C.~Amsler {\it et al.}  [Particle Data Group], Phys.\ Lett.\  B {\bf 667}, 1 (2008).



\bibitem{Cheng&Yang_PoleModel}
H.~Y.~Cheng and K.~C.~Yang, Phys.\ Rev.\  D {\bf 66}, 014020 (2002).

\bibitem{wfswave}
W.~Loinaz and R.~Akhoury, Phys.\ Rev.\  D {\bf 53}, 1416 (1996);
C.~H.~Chou, H.~H.~Shih, S.~C.~Lee and H.~n.~Li, Phys.\ Rev.\  D {\bf 65}, 074030 (2002).

\bibitem{wavefnB} F. Schlumpf, hep-ph/9211225.
\bibitem{wavefnBcBc} Y.Y. Keum, H.-n. Li, and A.I. Sanda, Phys. Lett. B {\bf 504}, 6 (2001).

\bibitem{Bcdecay}
H.~H.~Shih, S.~C.~Lee and H.~n.~Li, Phys.\ Rev.\  D {\bf 61}, 114002 (2000).

\bibitem{Dstopn}
C.~H.~Chen, H.~Y.~Cheng and Y.~K.~Hsiao, Phys.\ Lett.\  B {\bf 663}, 326 (2008).

\bibitem{decaywidth}
C.~Chen, X.~L.~Chen, X.~Liu, W.~Z.~Deng and S.~L.~Zhu, Phys.\ Rev.\  D {\bf 75}, 094017 (2007).


\bibitem{Dwave}
H.~Y.~Cheng and C.~K.~Chua, Phys.\ Rev.\  D {\bf 75}, 014006 (2007).

\bibitem{Korner94} J.G. K\"orner, M. Kr\"amer, and D. Pirjol, Prog.
Part. Nucl. Phys. {\bf 33}, 787 (1994).

\bibitem{CCH}
 H.~Y.~Cheng, C.~K.~Chua and C.~W.~Hwang,
 Phys.\ Rev.\ D {\bf 69}, 074025 (2004). 

\bibitem{pentaquark}
  H.~Y.~Cheng, C.~K.~Chua and C.~W.~Hwang,
  Phys.\ Rev.\  D {\bf 70}, 034007 (2004).

 \bibitem{Cheng97} H. Y. Cheng, C. Y. Cheung, and C. W. Hwang, Phys.
        Rev. D {\bf 55}, 1559 (1997).

 \bibitem{Jaus90} W. Jaus, Phys. Rev. D {\bf 41}, 3394 (1990).

 \bibitem{deAraujo:1999cr}
            W.~R.~de Araujo, M.~Beyer, T.~Frederico, and H.~J.~Weber,
            J.\ Phys.\ G {\bf 25}, 1589 (1999). 

  \bibitem{Brodsky:1997de}
  S.~J.~Brodsky, H.~C.~Pauli, and S.~S.~Pinsky,
  Phys.\ Rept.\  {\bf 301}, 299 (1998).




\end{thebibliography}
\end{document}